# Machine Learning-Based Analysis of Critical Process Parameters Influencing Product Quality Defects: A Real-World Case Study in Manufacturing

Sukumaran Rajasekaran[a], Ebru Turanoglu Bekar[*,a], Kanika Gandhi[b], Sabino Francesco Roselli[c], Mohan Rajashekarappa[a]

[a]*Department of Industrial and Materials Science, Chalmers University of Technology, SE-412 96 Gothenburg, Sweden*
[b]*Volvo Trucks Operation, Skövde, Sweden*
[c]*Department of Electrical Engineering, Chalmers University of Technology, SE-412 96 Gothenburg, Sweden*

**Abstract**

Quality control is an essential operation in manufacturing, ensuring products meet the necessary standards of quality, safety, and reliability. Traditional methods, such as visual inspections, measurements, and statistical techniques, help meet these standards but are often time-consuming, costly, and reactive. With the advent of AI/ML, manufacturers can shift from reactive to proactive approaches in quality control. This study applies ML-based models for predictive quality control in a real-world manufacturing setting. The case company produces castings for powertrain components in heavy vehicles, where poor control of core-making process parameters leads to costly defects. ML models were developed by analyzing data from two core-making machines, their processes, and maintenance logs to identify parameters associated with casting defects, enabling the prediction and prevention of potential defects before they occur. The results demonstrated good accuracy rates, helping quality and production teams identify and eliminate defective cores and thereby improving product quality and production efficiency.

## 1. Introduction

Quality control is a critical operation of manufacturing, ensuring that products meet essential standards for quality, safety, and reliability. Without proper quality control processes, manufacturers risk producing defective or unsafe products, leading to potential recalls, legal issues, and damage to their brand reputation [1]. In traditional quality control, products are inspected and tested at various stages of production to verify they meet defined specifications. Common methods include visual inspections, measurements, functional tests, and statistical process control (SPC) [2], [3]. While effective, traditional quality control can be time-consuming, expensive, and reactive since it identifies defects after they occur rather than preventing them in the first place [4].

The rise of new technologies and digitalization in manufacturing addresses these drawbacks by enabling real-time monitoring, predictive maintenance, and automated quality control [5]. Artificial Intelligence (AI) and Machine Learning (ML) can predict potential quality issues before they arise, allowing for early intervention [6]. The aim of this study is to showcase ML-based predictive quality control application in a real-world manufacturing setting. The goal is to develop a data-driven approach using ML to predict and prevent quality defects in the core-making process from a foundry line of the case company. This study contributes to the advancement of data-driven quality control in manufacturing by identifying the primary quality issues in a foundry line, analyzing the influence of critical process parameters and maintenance actions on casting defects, and applying suitable ML models to predict these defects. By integrating multiple data sources from process, maintenance logs, and quality records, the study demonstrates how ML can be effectively used to support proactive decision-making and improve product quality and production efficiency in a real-world manufacturing setting.

The rest of this paper is organized as follows: Section 2 presents the theoretical background by providing insights into casting processes, common casting defects, and ML-based quality control applications in manufacturing. Section 3 outlines the inspired methodology used to achieve the aim of the study. The results are detailed in Section 4, followed

---

* Corresponding author
 *E-mail addresses:* `rsukumar97@gmail.com` (Sukumaran Rajasekaran), `ebrut@chalmers.se` (Ebru Turanoglu Bekar), `kanika.gandhi@volvo.com` (Kanika Gandhi), `rsabino@chalmers.se` (Sabino Francesco Roselli), `rmohan@chalmers.se` (Mohan Rajashekarappa)

by a discussion and potential implementation strategies for industry in Section 5. Finally, Section 6 concludes the study by summarizing the key findings.

## 2. Theoretical Background

This section provides a detailed overview of key terms and concepts related to foundry, casting process and quality defects, establishing the foundational knowledge necessary to understand the case study presented in this paper. It also includes a summary review of existing research on ML-based approaches in manufacturing, with a focus on their application in quality control for foundry.

*2.1. Casting processes and defects in foundry*

The foundry industry forms the backbone of modern manufacturing by providing a cost-effective and versatile method of producing complex metal parts. In a foundry, molten metal is poured into a mold cavity that mirrors the desired shape of the final component. Once the metal solidifies, the mold is broken apart, and the casting is removed. This process, known as casting, is used extensively in sectors such as automotive, aerospace, energy, and heavy machinery for parts requiring intricate shapes and durability [7].

The core-making process, a critical phase within the foundry workflow, uses sand, resin, and hardeners to produce internal features of cast parts. A well-prepared core ensures dimensional accuracy and surface finish, directly affecting the quality of the final component. Various binder systems such as cold box, hot box, and shell processes are employed depending on the desired core characteristics, production volume, and complexity [8]. Parameters such as sand temperature, binder ratio, gassing time, and curing conditions need tight control to avoid internal casting defects like blowholes or incomplete cavities [9].

Casting defects are unavoidable in metal casting processes due to their dependence on multiple physical, chemical, and thermal phenomena. These defects can be broadly classified into categories such as surface defects (e.g., roughness, laps), internal defects (e.g., porosity, blowholes), dimensional defects, and metallurgical defects (e.g., segregation) [10]. Common defect types include blowholes, shrinkage cavities, cold shuts, hot tears, sand inclusions, and misruns. The root causes range from improper mold filling and poor core strength to gas entrapment and rapid solidification [10]. Most defects originate from poor sand quality, insufficient venting, inadequate maintenance, or unmonitored variation in process parameters [11]. Minor improvements in mold design or curing time can reduce the incidence of defects by over 40% in high-volume foundry operations [11].

To ensure the consistent quality of cast parts, traditional quality control systems have relied on inspection techniques such as visual testing, ultrasonic testing, magnetic particle inspection, and radiographic analysis. These post-process techniques can detect surface and internal defects but often lead to high scrap rates and additional costs, as the defect is only discovered after production [12]. Preventive approaches, such as Failure Mode and Effects Analysis (FMEA), SPC, and Six Sigma, have been implemented in some foundries to enhance process control and defect reduction [13]. However, traditional quality systems suffer from significant limitations. They are typically reactive rather than proactive, require manual labor, are time-consuming, and rely heavily on operator experience and judgment [2-4]. Therefore, in today's highly competitive market, relying solely on these traditional quality control approaches leads to inefficiencies and increased costs. Additionally, the complex interdependencies among variables in casting processes make them difficult to monitor and optimize using these approaches. This highlights the need for more advanced and data-driven approaches, enabling manufacturers to transition from reactive to proactive quality control strategies [14].

*2.2. Related work on ML-based approaches for quality control in foundry*

The arrival of Industry 4.0 technologies, including cyber-physical systems, Industrial Internet of Things (IIoT), and Big Data Analytics, has been transforming manufacturing operations into smart factories. In these environments, real-time data collected from machines, sensors, and systems is analyzed to detect anomalies, predict failures, and improve process stability [4], [5]. ML, a key pillar of AI, has been emerging as a transformative technology in the manufacturing industry [14], [15]. ML models learn from historical data to identify patterns, correlations, and anomalies, enabling accurate predictions and automation of complex decision-making processes.

In foundry operations, the integration of such smart technologies is particularly crucial due to the high complexity and numerous interdependent variables involved. Advanced sensor systems continuously track key core-making conditions such as sand temperature, humidity, air pressure, and gas curing levels [16], [17]. The data collected from these sensors serve as inputs for ML models, which are capable of identifying defect patterns even before the casting process begins, often by detecting anomalies in core production or binder curing stages [6]. By enabling real-time analysis of process parameters, ML facilitates defect tracking and prediction, thereby shifting quality control strategies from traditional reactive methods to proactive, predictive defect prevention [6], [14].

Supervised ML algorithms, such as decision trees, random forests, and gradient boosting classifiers, are among the most commonly applied techniques in manufacturing quality control [15]. These models are particularly effective in binary or multi-class classification problems such as distinguishing between defective and non-defective parts [18], [19]. Decision tree algorithms work by recursively splitting the dataset based on the most influential variables, forming a tree structure where each node represents a decision rule. While intuitive and easy to interpret, single decision trees often suffer from overfitting and limited generalizability [20]. To overcome this, ensemble methods like random forests and gradient boosting have been developed. Random forests construct multiple decision trees from bootstrapped subsets of the data, combining their predictions through voting. This approach reduces variance and enhances model robustness, making it ideal for high-dimensional, noisy manufacturing datasets. Gradient boosting, in contrast, builds trees sequentially, where each new tree corrects errors made by the previous one, achieving high accuracy with careful tuning [18], [20].

Recent studies have demonstrated the feasibility of ML in defect detection and root cause analysis. For example, BramahHazela et al. [21] used supervised algorithms in high-precision foundry operations and achieved significant improvements in early defect detection. One challenge foundries face in implementing ML is the class imbalance between defective and non-defective parts. Most components are typically defect-free, making models biased toward the majority class. Techniques such as undersampling, oversampling, and synthetic minority over-sampling technique (SMOTE) help mitigate this issue, improving prediction reliability [22]. Another challenge is feature selection and engineering. In foundry settings, hundreds of process parameters may be captured, ranging from sand particle size and mold wall thickness to curing time and machine calibration settings. Feature importance scores from tree-based models can identify which parameters most influence casting quality, enabling targeted process improvements [21]. To address the aforementioned challenges as in implementing ML in industrial settings, this study adopts the Cross-Industry Standard Process for Data Mining (CRISP-DM) methodology, which is widely recognized in industrial ML applications [23]. Applying such methodology can help align quality control practices with predictive analytics efforts, ensuring more structured and effective outcomes [24-28].

## 3. Methodology

To establish a structured approach for ML-based analysis of critical process parameters affecting product quality defects, the CRISP-DM is followed as a reference methodology in this paper. CRISP-DM is domain-agnostic process model that supports iterative development and deployment of ML solutions. It offers a systematic approach including six phases such as business understanding, data preparation, modeling, evaluation, and deployment as demonstrated in Figure 1.

The *business understanding phase* involved identifying the main quality challenges faced by the foundry, such as recurring casting defects and their cost implications. In the *data understanding phase*, historical data from core-making machines, maintenance logs, and defect reports were collected and examined to assess data quality and relevance. The *data preparation phase* focused on cleaning, transforming, and integrating this data to create a robust dataset suitable for modeling. During the *modeling phase*, supervised ML algorithms were applied to identify patterns and correlations between process parameters, maintenance actions, and casting defects. The *evaluation phase* assessed model performance using domain-relevant metrics to ensure its reliability for decision-making. Evaluation metrics are essential to assess ML model performance and guide model selection [29]. Finally, the *deployment phase* outlines how these insights can be integrated into production workflows to support proactive quality control and aid operators in identifying potentially defective cores before casting.

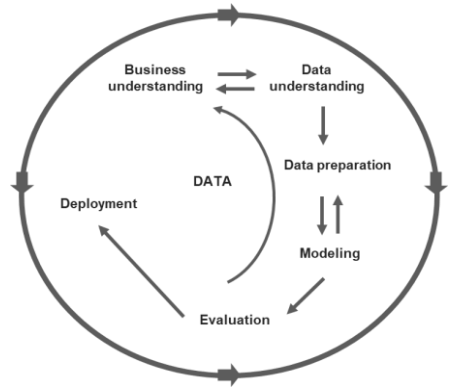

Figure 1: CRISP-DM methodology adapted from [23].

## 4. Results

This section presents the results corresponding steps of the CRISP-DM methodology outlined in Section 3. During the implementation of some steps (e.g., data understanding, data preparation, and modeling), the Python programming language carried out using libraries such as Matplotlib and Seaborn for data visualization, NumPy and Pandas for data preparation, and Scikit-learn for ML modeling.

*4.1. Business understanding*

The business objective was to develop a data-driven approach using ML to predict and prevent quality defects in foundry line at the case company, which manufactures powertrain components. The focus area was the sand casting process, with its detailed workflow illustrated in Figure 2.

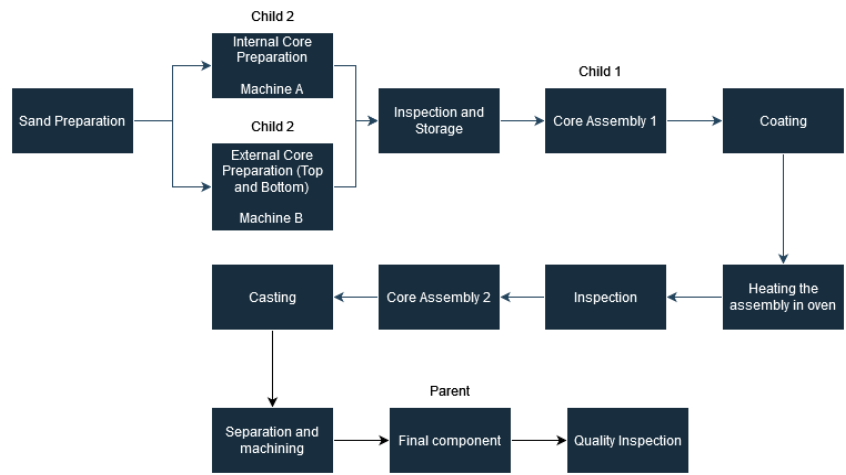

Figure 2: Flowchart of the casting process in the foundry line.

The process begins with sand preparation, where sand is mixed with binders and additives to achieve the desired properties. This mixture is then conveyed to Machine A and Machine B, responsible for forming internal and external cores, respectively. These cores undergo inspection before proceeding to core assembly 1, where robots position the internal core into the bottom core using camera-assisted alignment. The assembled cores are then coated and heated in an oven to improve surface finish, thermal insulation, and reduce core-metal interaction. A manual inspection follows to ensure core integrity. In core assembly 2, the top core is added, and the complete assembly is sent to the casting bay. Molten metal is poured into the assembled cores, which are then cooled and separated. After removing excess material through machining, a final inspection ensures the casting meets quality standards.

*4.2. Data understanding*

In this study, four different datasets from multiple sources relevant business objective were selected to support the analysis and development of ML models. The first dataset consists of *product data from the Manufacturing Execution System (MES)* [30], which encompasses all the components generated in the foundry line. In order to cast the component, several elements are necessary: a top core, a bottom core that maintains the outer shape of the component, and a series of internal cores positioned between these two halves. These internal cores bring forth the intricate internal structures required for the component. This data has 9 attributes including information about each individual core and the sequential operations performed on them with the timestamps for the various operations conducted on these cores. The second dataset, *quality deviation data, is sourced from the foundry's Quality System* managed by the Quality Control Division. It includes 26 attributes detailing defective components, such as serial number, component name, deviation type, specific production line where the deviation occurred, cost incurred by the defective component, origin of the deviation (problem owner), casting date of the product, and registration date of quality deviation in the casting. This dataset supports nonconformity tracking, root cause analysis, and process improvement. The third dataset, *process data extracted from the MES*, focuses on two pilot machines selected for this study, referred to as Machine A and Machine B. These machines are responsible for producing internal and external cores. The Programmable Logic Controller (PLC) system integrated into Machine A and Machine B systematically logs every process step carried out within these machines. These logged data from both machines are vital features for the analysis, aiming to investigate how machine operations influence final product quality. The dataset includes 11 attributes such as operation codes, descriptions, timestamps, product IDs, actual vs. target values, and tolerances. The final dataset includes *maintenance records for Machines A and B, obtained from the Computerized Maintenance Management System (CMMS)* [30]. This dataset encompasses relevant information, including breakdown frequency, types of failures, timestamps of each occurrence, and total downtime. It provides valuable insights into how machine reliability may influence product quality and overall operational efficiency within the foundry line.

During the data exploration phase, visualization techniques were used to understand patterns within datasets comprising 571 673 products manufactured between a specific period (January 2020 and May 2023). This phase aimed to enhance data familiarity and identify key quality issues. It was found that approximately 54% of quality deviations originated from the casting process and 38% from core making, together accounting for over 90% of all defects. Further analysis revealed that five specific defect types, e.g., pore holes, wrongly assembled cores, blow holes, core bits, and disintegration were responsible for 88% of total deviations as shown in Figure 3. Only these top five were selected for further analysis. In terms of maintenance activity, 350 operations were recorded for Machine A and 210 for Machine B during the same period.

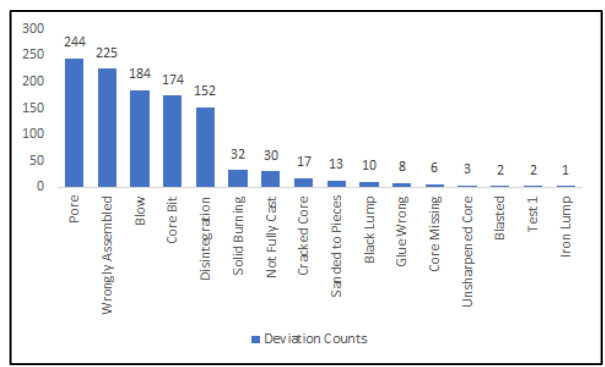

Figure 3: Distribution of deviation across different processes.

*4.3. Data preparation*

Key attributes from the quality deviation dataset such as deviation types, problem owners, and defect classifications) were first translated for better understanding. The next step involved linking quality deviation data with MES product data to trace which cores contributed to defective components. This task was complex due to the hierarchical structure in the MES system, where second- and third-level child relationships were not explicitly defined. Multiple iterations were performed to trace the lowest-level core components produced by Machines A and

B. Once identified, this core-level data was separately merged with the corresponding process data from each machine to evaluate how machine-specific operations influenced final product quality. Pivot tables were created for both machines to restructure the process descriptions into feature columns. These were further merged with maintenance data to examine the effect of machine breakdowns on defect occurrence.

During data preparation, several key steps were implemented to ensure data quality and consistency for model training. This included imputing missing values using median imputation, removing noisy records based on logical constraints, applying one-hot encoding for categorical variables, and normalizing numerical features using Min-Max scaling [31].

Identifying the most influential features for each of the top five defect categories as shown in Figure 3, e.g., "Pore," "Wrongly Assembled," "Blow," "Core Bit," and "Disintegration" was one of key objective during the analysis. However, the final dataset was highly imbalanced, with the "None" class (representing defect-free cores) overwhelmingly dominating. This imbalance stems from the fact that most cores produced do not exhibit defects, with only a small portion associated with deviations. The presence of class imbalance poses a challenge in constructing an unbiased and effective ML model. A model trained on imbalanced data may be biased towards the majority class, leading to sub optimal performance on the minority classes. To address this class imbalance problem, our merged dataset was rebalanced by undersampling the majority class [22] to ensure equal representation of defective and non-defective samples and to maintain simplicity and control model bias. It should be also noted that there are other techniques such as SMOTE or cost-sensitive learning can also be used to address class imbalance problem. The resulting dataset was clean, balanced, and ready for use in ML modeling.

*4.4. ML modeling*

Two supervised ML models, Random Forest and Gradient Boosting, were applied as binary classifier to the cleaned and balanced dataset for predicting core quality defects in the foundry line. These models were selected for their effectiveness in handling complex, non-linear relationships and their capability and explanability to identify and rank important features influencing quality outcomes. Random Forest is an ensemble learning method that constructs multiple decision trees using random subsets of data and features [18], [32]. It aggregates the predictions of these trees to enhance overall accuracy, making it well-suited for datasets with a large number of features [18], [32]. Gradient Boosting, on the other hand, builds models sequentially, with each new model aiming to correct the errors of the previous one. This approach is highly effective for both classification and regression tasks across various domains. For a deeper understanding of decision trees, ensemble methods like Random Forest and Gradient Boosting, including their mechanisms, strengths, and use cases in ML, readers are encouraged to read reference [33]. It should be also noted that the data was split into 80% for training and 20% for testing and evaluation, and default parameters in Scikit-learn has been used during the modeling. The objective was to identify key features contributing to the five most common defect types and show the predictive performance for the certain defect type.

*4.5. Evaluation and deployment*

To assess model performance comprehensively, four evaluation metrics were used: accuracy, precision, recall, and the F1-score. Accuracy measures the proportion of total correct predictions, while precision indicates the proportion of true positives among all predicted positives which highlights how reliable the model is when it finds a defect. Recall measures the model's ability to correctly identify all actual defects, which is particularly important in quality control. The F1-score, which balances both precision and recall, is especially useful in imbalanced datasets, ensuring neither metric dominates. A more detailed explanation of these metrics and their relevance to classification problems in manufacturing can be found in [29]. The evaluation metrics confirmed that both ML models performed closely and effectively across all key metrics. For a simple comparison, accuracy was used as the evaluation metric. On Machine A, Random Forest achieved an avg. accuracy of 66.2%, while Gradient Boosting performed at avg. 62.4%. On Machine B, Random Forest achieved avg. 56% accuracy, while Gradient Boosting scored avg. 54%. The models also displayed different behavior in false positive and false negative predictions. The results are detailed in Table 1.

While standard evaluation metrics such as accuracy, precision, recall, and F1-score were used to assess model performance, it is important to interpret these metrics in the context of real-world manufacturing priorities. Specifically, recall should be prioritized when the goal is to detect as many actual defects as possible, reducing the risk of false negatives, which can lead to defective products reaching customers. On the other hand, precision reflects the model's ability to avoid false positives, helping prevent unnecessary inspections or rework. F1-score balances both metrics, offering a useful single measure when there is a trade-off between defect detection and operational efficiency. Accuracy, while informative, may be misleading in imbalanced datasets and should be interpreted with caution. In quality control, recall is particularly crucial to ensure that actual defects are not overlooked (i.e., minimizing false negatives), while precision helps avoid excessive false alarms and unnecessary interventions. In practice, the cost of missing a defect (false negative) is often much higher than the cost of falsely flagging a good part (false positive). Therefore, future implementations could benefit from a cost-sensitive evaluation framework that can help translate model performance into meaningful business impact.

Table 1: Comparison of Random Forest (RF) and Gradient Boosting (GB) models on Machine A and Machine B across each defect type.

| Defect type | ML model | Accuracy for Machine A | Accuracy for Machine B |
|---|---|---|---|
| Pore | RF | 0.63 | 0.41 |
| | GB | 0.61 | 0.41 |
| Wrongly Assembled | RF | 0.65 | 0.60 |
| | GB | 0.63 | 0.58 |
| Blow | RF | 0.64 | 0.64 |
| | GB | 0.68 | 0.62 |
| Core Bit | RF | 0.74 | 0.59 |
| | GB | 0.63 | 0.51 |
| Disintegration | RF | 0.65 | 0.56 |
| | GB | 0.57 | 0.56 |

The study does not include the deployment phase but instead provides recommendations to the company, emphasizing the importance of robust data collection practices to enhance data quality and prediction accuracy for future applications. Additionally, the proof of concept demonstrated by the ML models not only raises awareness of their potential but also illustrates the company's readiness regarding organizational and IT infrastructure to adopt and implement these solutions within the foundry.

## 5. Discussion

This study applied two supervised ML models which were Random Forest and Gradient Boosting to predict core quality defects in a foundry production line. These models were chosen for their ability to manage high-dimensional data and capture non-linear patterns. Random Forest, known for its robustness and ease of interpretation, aggregates predictions from multiple decision trees. Gradient Boosting, by contrast, builds models sequentially to correct previous errors, offering enhanced predictive performance. Across datasets from both Machine A and Machine B, Random Forest consistently performed slightly better than Gradient Boosting in prediction accuracy. The results suggest that model selection should consider not only accuracy but also the data characteristics and variability across machines.

Feature importance was derived using the Gini importance (mean decrease in impurity) for the Random Forest model and the feature contribution scores from the trained Gradient Boosting model. These scores were extracted using the Scikit-learn library's attribute. These values were normalized and visualized for both Machine A and Machine B to interpret the key drivers of each defect type. Feature importance analysis showed that sand temperature was the most influential parameter across both models. Other important features included binder quantity, gassing time, and core box temperature, all of which strongly correlated with defect formation. Notably, Machine B exhibited greater sensitivity to factors such as maintenance intervals and cycle duration, which had less impact on Machine A. This distinction emphasizes the variability in defect causation between different machines and supports the need for machine-specific quality strategies. The feature importance scores of each machine for each defect type are presented in Figure 4 and Figure 5, respectively. Based on the findings, several recommendations can be proposed due to the potential of ML models to improve core-making processes and predict product quality. Their integration can transform manufacturing by providing proactive, data-driven insights [14]. Integrating the models into a real-time MES could enable early warnings and preventive interventions on the shop floor. A hybrid approach, combining sensor data with process logic and expert input, could further improve model accuracy. Exploring deep learning models like LSTM could also enhance the system's ability to capture time-dependent patterns in sensor data. Expanding the scope to include casting outcomes and environmental data would offer a more complete view of the process taking into account the sustainability perspective. Additionally, implementing a retraining pipeline would help keep the models updated with evolving production conditions. This recommendation highlights promising avenues for future research. However, as now clarified, these approaches can also present challenges including the need for continuous sequential data and effective sensor integration. Addressing these challenges will require careful consideration of data availability and methodological requirements, underscoring the complexity involved in advancing model performance beyond the current scope.

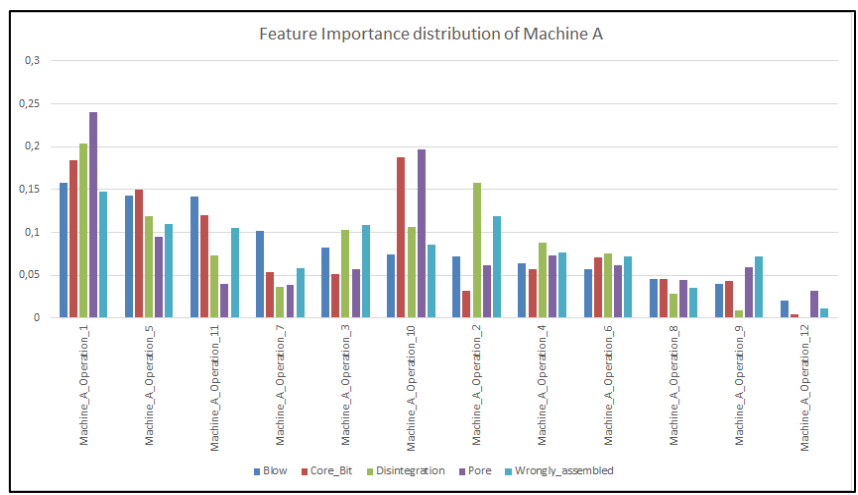

Figure 4: Feature importance scores of Machine A.

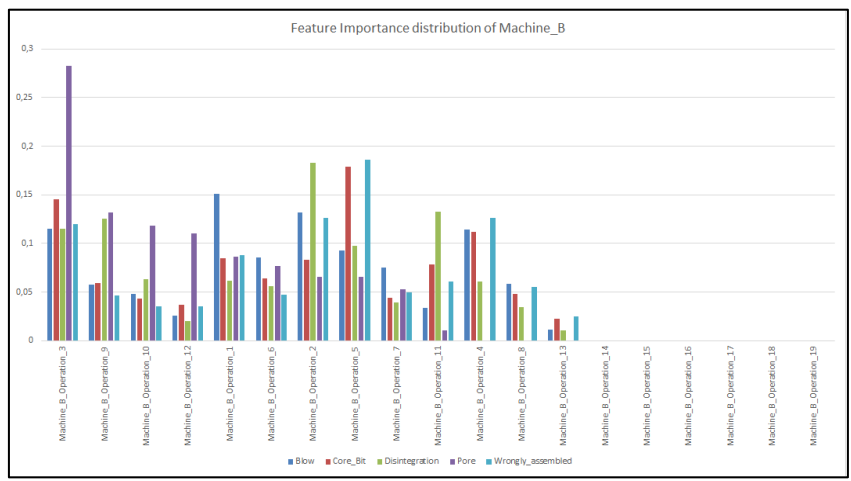

Figure 5: Feature importance scores of Machine B.

As a summary, this study contributes to both academic and industrial domains. Academically, it illustrates how ensemble ML learning models, e.g., Random Forest and Gradient Boosting, can be effectively applied in real-world manufacturing settings using the CRISP-DM methodology. Industrially, it provides a proof-of-concept for predictive quality control using and combining data from multiple sources including production, quality, and maintenance systems. It highlights critical process parameters affecting core quality and outlines a clear methodology for building, evaluating, and potentially deploying ML models in production environments

## 6. Conclusion

In conclusion, this study highlights the significant potential of ML models in analyzing critical process parameters and identifying root causes of specific casting defects within a foundry environment. The findings show the value of ML as a powerful analytical tool capable of uncovering complex patterns and insights in manufacturing data. By applying the CRISP-DM methodology, the study demonstrates a structured and practical approach for implementing data-driven solutions in industrial contexts. This approach not only enhances process understanding but also supports more informed decision-making for operators in proactively addressing quality issues, thereby enhancing production efficiency and quality assurance, reinforcing the broader impact of this research. Empirical results further validate the practical utility of the models, with the ML approach achieving an acceptable accuracy in detecting defective cores. While the developed models showed promising results, there remains room for improving predictive accuracy, which can be the focus for further research. Additionally, real-world implementation may present challenges related to system integration, data quality, and organizational readiness, which require careful planning and adaptation of ML. In summary, this work contributes meaningfully to the field of quality control by showcasing how AI/ML can drive smarter, more efficient manufacturing. These technologies hold the potential to optimize resource use, reduce operational costs, and support for advancing toward data-driven, intelligent manufacturing systems.

**Acknowledgements**

This paper is based on the Master's thesis conducted by the first author at Chalmers University of Technology. The authors gratefully acknowledge the support of the Advanced and Innovative Digitalization Program, funded by VINNOVA, through the research project TPdM - Trustworthy Predictive Maintenance (Grant No. 2022-01710). The study was carried out within the Production Area of Advance at Chalmers University of Technology. The authors also extend their appreciation to the personnel from the Maintenance, Production, and Quality departments of the case company for their valuable support and assistance with data and domain expertise.